\documentclass[aps,prb,twocolumn]{revtex4-1}

\usepackage{dcolumn,amsfonts,amsmath,amssymb}
\usepackage{bm}
\usepackage{lipsum}
\usepackage{graphicx}

\begin{document}
\title{Inter-band Coulomb coupling in narrow gap semiconductor nanocrystals: \mbox{$\bm{k}\cdot\bm{p}$} theory}
\author{Maryam Azizi}
\author{Pawe{\l} Machnikowski}
\affiliation{Institute of Physics, Wroc{\l}aw University of Technology, 50-370 Wroc{\l}aw, Poland}
\date{\today}

\begin{abstract}
\label{abstract}
We derive the matrix elements of Coulomb interaction between states with 
different number of electrons and holes in a semiconductor nanocrystal 
within the 8-band \mbox{$\bm{k}\cdot\bm{p}$} theory. These matrix elements are responsible 
for multiple exciton generation which may contribute to the enhancement 
of the efficiency of solar cells. Our calculations are performed within the 
multi band envelope function formalism based on the states resulting 
from diagonalization of the 8-band \mbox{$\bm{k}\cdot\bm{p}$} Hamiltonian. 
We study in detail and compare two contributions to the inter-band 
Coulomb coupling: the mesoscopic one, which involves only the envelope 
functions and relies on band mixing, and the microscopic one, that relies 
on the Bloch parts of the wave functions and is non-zero even between single-
band states. We show that these two contributions are of a similar order of 
magnitude. We study also the statistical distribution of the
magnitudes of the inter-band Coulomb matrix elements and show that the
overall coupling to remote states decays according to a power law
favorable for the convergence of numerical computations.
\end{abstract}

\maketitle

\section{Introduction}
\label{introduction}
Semiconductor nanocrystals (NCs) are of considerable current interest for 
exploring a large number of novel phenomena at the nanoscale and for exploiting 
their unique size dependent properties in potential applications. 
In particular, these semiconductor nanostructures have a large potential
for applications in nano- and optoelectronics\cite{talapin10,teperik12,LinPe,binks11}.

\begin{figure}[tb]
\centering
\includegraphics[width=85mm]{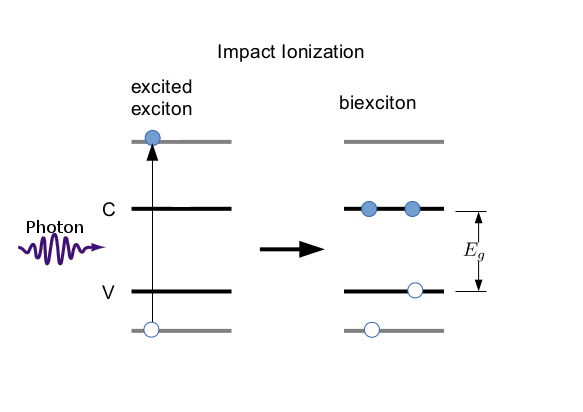}
\caption{\label{states}
(Color online) Schematic depiction of multiple exciton generation by impact ionization of a 
high-energy electron-hole pair.
} 
\end{figure}

One of the interesting properties of semiconductor nanocrystals is 
the interband Coulomb coupling \cite{deuk11,deuk12,franceschetti06,rabani08,califano09,
baer12,allan06,delerue10,shabaev06,witzel10,Silvestri,korkusinski10} 
that can lead to multiple exciton generation (MEG)\cite{binks11,sambur10,semonin11}. 
In the MEG process, absorption 
of a single photon leads to creation of two or more electron-hole pairs, 
as schematically depicted in Fig.~\ref{states}. This can occur when absorbing a photon 
is followed by creation of an electron-hole pair (an exciton) which then relaxes into an 
energetically lower state and the excess energy is used to create a second 
electron-hole pair (thus creating a biexciton
state)\cite{schulze11,maryam,franceschetti06,rabani08,califano09, 
baer12,allan06}.  
The same process can also occur coherently via a 
superposition of single- and bi-exciton states, or with the single exciton
state playing the role of a virtual intermediate state\cite{schaller05,rupasov07,shabaev06,witzel10,Silvestri,kowalski09}. 
In any case, this process is mediated by Coulomb scattering between electron states 
in different bands which does not conserve the number of electron-hole pairs.
Experiments indicate that the MEG process may indeed
contribute to the efficiency of solar cells\cite{sambur10,semonin11}.

Because of its importance both for the full understanding of 
nanocrystal properties as well as for practical applications, 
the interband Coulomb couplings were extensively studied by 
various theoretical methods, including density functional theory \cite{deuk11,deuk12}, 
pseudopotential method\cite{franceschetti06,rabani08,califano09,baer12}, 
or tight binding approach \cite{allan06,delerue10,korkusinski10,korkusinski11}. 
Since the ab-initio and atomistic methods are computationally expensive, 
the less numerically demanding \mbox{$\bm{k}\cdot\bm{p}$} method has also been used in the 
modeling of Coulomb couplings \cite{shabaev06,witzel10,silvestri10,piryatinski10,Velizhanin}.
This approach allows one to perform more extensive computations including 
coupled exciton and biexcition states in a wide energy range. 

When viewed from the \mbox{$\bm{k}\cdot\bm{p}$} perspective, the Coulomb coupling between 
few-particle states with different numbers of electron-hole pairs can 
appear in two ways. First, it can be due to band-mixing, with the 
two states coupled by the usual intraband Coulomb interaction terms 
involving, e.g., the conduction band admixture to a hole state \cite{Velizhanin}. 
We will refer to this contribution as ``mesoscopic" as the relevant matrix 
element involves only the envelope parts of the wave functions 
(similarly to the common electron-electron or electron-hole interactions 
computed in the usual way in a nanostructure). Second, the coupling between 
such configurations can appear directly when one takes the Bloch part of 
the wave function into account \cite{kowalski09}. This is, in turn, formally similar 
to the \mbox{$\bm{k}\cdot\bm{p}$} calculation of electron-hole exchange coupling in a quantum 
dot \cite{Kadantsev}. 
Because of its formal structure, we will refer to this contribution as 
``microscopic". This microscopic contribution is formally reduced by a 
factor on the order of $a/R$ as compared to the mesoscopic one \cite{kowalski09} 
(where $a$ is the lattice constant and $R$ is the nanocrystal radius). 
In the absence of band mixing, the leading order term in the expansion 
of the Coulomb coupling is proportional to $a/R$. If band mixing is present, 
a contribution on the order of 1 will appear. Such terms are, by themselves, 
two orders of magnitude stronger than the first-order ones but their contribution 
is greatly reduced due to small amount of valence band admixture to conduction 
band states (or vice versa). On the other hand, since the former (microscopic) 
term does not rely on band mixing, it is not possible to predict \textit{a priori} whether it will be 
small compared to the mesoscopic one. 
In any case, keeping terms up to the first order in $a/R$ in the microscopic 
part and the zeroth order in the mesoscopic part is sufficient to capture both 
contributions to the leading order.
Estimates obtained using a simplified, 
single-band model of wave functions \cite{kowalski09} yield values of the 
microscopic part of the matrix element up to several meV or a 
few tens of meV and the resulting degree of mixing between 
single-exciton and biexciton states is on the order of 0.1, which 
suggests that this contribution is not negligible.

In this paper, we present the calculations of the Coulomb
matrix elements between exciton (X) and biexciton (BX) states within the 
multi band envelope function formalism based on the states resulting 
from diagonalization of the 8-band \mbox{$\bm{k}\cdot\bm{p}$} Hamiltonian\cite{efros98}. Mesoscopic 
and microscopic contributions to the inter-band Coulomb coupling 
are studied in detail by generalizing the previous results \cite{kowalski09} to the 
realistic model of wave functions including the band mixing.
The relatively low computational cost of the \mbox{$\bm{k}\cdot\bm{p}$} method allows us to 
find coupled pairs of X-BX states in a
very broad energy window and to study the distribution of
the magnitudes of the matrix elements vs. the energies of
the coupled configurations in order to build reliable statistics. 
We show that in many cases the two contributions are 
of a similar order of magnitude so both need to be taken into account
for reliable modeling. Moreover, we analyze the statistics of the
coupling magnitudes relative to the energy distance between the two
coupled states which allows us to assess the contribution of remote
states to the X-BX state mixing. We confirm the findings of the simplified model\cite{kowalski09} 
and show that this contribution
decreases with energy distance, thus providing formal grounds for
restricting numerical computations to a finite energy window. 

The paper is organized as follows. In Sec.~\ref{sec:model}, we define
the model. In Sec.~\ref{sec:Carrier states and matrix elements} we
discuss the derivation and computation of the inter-band coulomb matrix elements.
In Sec.~\ref{sec:results}, the results of our calculations are presented are discussed. 
Finally, Sec.~\ref{sec:concl} concludes the paper.

\section{Model}
\label{sec:model}
In this section, we describe the model of the nanocrystal used 
for our calculations.

We consider a simple model of a nanostructure, defined as an 
InAs sphere with an infinitely high potential barrier at its 
boundary. The radius of the nanostructure is $R=2.5$ nm. For the single-particle 
spectrum, we use the envelope function formalism with a standard 8-band 
Luttinger-Kohn Hamiltonian\cite{efros98}. All the material parameters relevant 
to the single-particle spectrum are taken from Ref.~[30].
The nanocrystal is assumed to be surrounded by air. 
We assume the dielectric constant of InAs $\epsilon_s = 14$.

The Coulomb energy in a spherical NC is composed 
of the direct Coulomb interaction and the coupling via surface polarization
due to dielectric discontinuity between the NC and the environment. The direct part is 
\begin{eqnarray}\label{directcoulomb}
 U_{\mathrm{direct}}(\bm{r},\bm{r}')=\frac{e^2}{4\pi\epsilon_{0}\epsilon_{s}}\frac{1}{|\bm{r}-\bm{r}'|}.
\end{eqnarray}
The indirect part contains the two-particle term describing mutual 
interaction of electrons via the polarization field,
\begin{eqnarray}\label{polcoulomb}
 \!\!\!\!U_{\mathrm{pol}}^{(2)}(\bm{r},\bm{r}')=-\frac{e^2}{4\pi\epsilon_{0}\epsilon_{s}}
\sum_k \chi_k \frac{(rr')^k}{R^{2k+1}}P_k\left(\frac{\bm{r}\cdot\bm{r}'}{rr'}\right),
\end{eqnarray}
and the single particle term
\begin{eqnarray}
 U_{\mathrm{pol}}^{(1)}(\bm{r})=\frac{e^2}{2\epsilon_2}\sum_{n=1}^N 
\sum_{k=0}^{\infty} \alpha_k \frac{r^{2k}}{R^{2k+1}},
\end{eqnarray}
which accounts for the self-energy contribution arising from the interaction 
of a charge with its own polarization field\cite{brus84}.
Here $P_k$ are Legendre polynomials and $\chi_k=(k+1)(\epsilon-1)/(k\epsilon_s+k+1)$.
In addition, interaction 
of the electrons with the positive ``jellium'' background yields further single-particle terms.

In view of the strong quantization of the energy levels in a 
small nanostructure we neglect Coulomb correlations and energy 
shifts for the few-particle configurations and include only 
the inter-band Coulomb couplings that are the essence of the present study.

\section{Carrier states and matrix elements}
\label{sec:Carrier states and matrix elements}
In this Section, we present the systematic derivation of the interband 
Coulomb matrix elements (that is, matrix elements coupling single- 
and bi-exciton states) within the 8-band envelope function approach. 
First, in Sec.~\ref{sec:Single particle states}, we present the single-particle states that make 
up the few-particle configurations. Then, in Sec.~\ref{sec:Interband Coulomb coupling}, we classify all 
the Coulomb terms of this kind and identify those relevant to the actual 
exciton-biexciton coupling. Finally, in Sec.~\ref{sec:Matrix elements}, we derive the matrix 
elements for the multi-band wave functions.

\subsection{Single particle states}
\label{sec:Single particle states}
Each electron and hole state is characterized by the total angular momentum 
$j$, the projection of the total momentum $-j\leq m\leq j$, the spatial inversion parity, 
and an additional quantum number $n$ labeling the subsequent wave functions with the same 
$j$, $m$ and parity.
We write the wave functions in the eight-band envelope approximation 
in the form
\begin{equation}\label{wavefunction}
 \Psi_{\gamma}^{\pm}(\bm{r},s)=\sum_{\lambda}\varphi_{\gamma\lambda}(\bm{r})u_{\lambda}(\bm{r},s),
\end{equation}
where $\bm{r}$ is position, $s$ denotes the spin projection, 
$\gamma$ stands for the set of quantum numbers ($jmn$), $\pm$ refers the parity, 
$\varphi_{\gamma\lambda}(\bm{r})$ is the envelope function 
and $u_{\lambda}(\bm{r},s)$ is lattice-periodic Bloch part. 
Here, $\lambda$ denotes the subband within the eight-band 
\mbox{$\bm{k}\cdot\bm{p}$} expansion: two subbands in the 
conduction band and six subbands in the valence band 
(heavy hole, light hole and spin-orbit split subbands).

In the numerical calculations, the envelope functions are further expanded into the basis 
functions composed of Bessel functions $j_l$ for the radial part and spherical harmonics 
$Y_{lm}$ for the angular dependence,
\begin{eqnarray}\label{envelope}
 \varphi_{\gamma\lambda}(\bm{r})& &=\sum_{nlm} c_{nlm}^{(\gamma\lambda)}N_{nl}j_l\left(\frac{r}{R}X_{ln}\right)Y_{lm}(\theta,\varphi)\nonumber\\
 & &\equiv\sum_{\nu}c_{\nu}^{(\gamma\lambda)}\psi_{\nu}(\bm{r}).
\end{eqnarray}
Here, $\nu$ represents the three quantum numbers, $nlm$ and $N_{nl}=\sqrt{2}/|j_{l+1}(X_{ln})|$. The summation in Eq.~\eqref{envelope} 
is over $l=j \pm 1/2$ or $l=j \pm 3/2$ depending on the subband $\lambda$\cite{efros98}. 
The coefficients $c_{\nu}^{(\gamma\lambda)}$ are found from numerical diagonalization of the
eight-band \mbox{$\bm{k}\cdot\bm{p}$} Hamiltonian\cite{efros98} with a cut-off for $n$ 
at $n_{\textrm{max}}=100$. The allowed optical transitions result from
the standard dipole selection rules with their relative magnitude
dependent on the overlap of the envelope functions. In particular, the
optical transitions are only allowed between conduction and valence  
band states with the same parity.

\subsection{Interband Coulomb coupling}
\label{sec:Interband Coulomb coupling}

In this section, we formally derive the general matrix elements 
between single-exciton and biexciton states (without any reference 
to the particular model of wave functions). We provide physical 
interpretation for the whole variety of these terms and indicate 
(on the grounds of the particular energetic relations for the case of interest) 
those relevant to the actual exciton-biexciton coupling.

In the ground state of a NC (to be denoted $|GS\rangle$), 
the valence band is fully occupied, while the conduction 
band is empty. Here, by valence (conduction) band we 
understand the single-particle eigenstates resulting 
from the \mbox{$\bm{k}\cdot\bm{p}$} diagonalization with energies below (above) 
the fundamental band gap (as opposed to the original bands 
of a bulk crystal at $k=0$). For the sake of more clarity in 
our derivations, the general index $\gamma$ in the expansion Eq.~\eqref{wavefunction} 
will be replaced by $\beta$ and $\alpha$ for the valence and 
conduction band states, respectively.
The corresponding creation (annihilation) operators are $a^{\dagger}_{\alpha}(a_{\alpha})$ 
and $a^{\dagger}_{\beta}(a_{\beta})$ in the conduction and valence
band, respectively. 
We denote the X and BX configurations by 
$|\alpha\beta\rangle = a_{\alpha}^{\dag}a_{\beta}|GS\rangle$ and 
$|\alpha_{1}\alpha_{2}\beta_{1}\beta_{2}\rangle = 
a_{\alpha_{1}}^{\dag}a_{\alpha_{2}}^{\dag}a_{\beta_{1}} a_{\beta_{2}}|GS\rangle$.

The single particle terms of the Hamiltonian (that arise from electron-ion 
interaction and the polarization self-energy) have the form 
\begin{eqnarray}
 H^{(1)}=\sum_{\gamma\gamma'}V_{\gamma\gamma'}^{(1)}a_{\gamma}^{\dagger}a_{\gamma'},
\end{eqnarray}
where $V_{\gamma\gamma'}^{(1)}=\sum_s \int d^3r \Psi_{\gamma}^{\star}(\bm{r},s) U^{(1)}(\bm{r}) \Psi_{\gamma'}(\bm{r},s)$. 
Here $U^{(1)}(\bm{r})$ denotes all the single particle terms in the Coulomb interaction.
The only non-zero contribution to the X-BX coupling is
\begin{eqnarray}
 \lefteqn{\langle\!\alpha_1\alpha_2\beta_1\beta_2|H^{(1)}|\alpha\beta\rangle} \nonumber\\
 &=&\sum_{\alpha'\beta'}V_{\alpha'\beta'}^{(1)}
 \langle\!GS|
 a_{\beta2}^{\dagger} a_{\beta1}^{\dagger} a_{\alpha2} a_{\alpha1} 
 a_{\beta'}^{\dagger} a_{\alpha'} a_{\alpha}^{\dagger} a_{\beta}|GS\rangle\nonumber\\
 &= &-V_{\alpha_1\beta_1}\delta_{\alpha_2\alpha}\delta_{\beta_2\beta}
 +V_{\alpha_1\beta_2}\delta_{\alpha_2\alpha}\delta_{\beta_1\beta}\nonumber\\
 & &+V_{\alpha_2\beta_1}\delta_{\alpha_1\alpha}\delta_{\beta_2\beta}-V_{\alpha_2\beta_2}\delta_{\alpha_1\alpha}\delta_{\beta_1\beta}.
\end{eqnarray}
All these terms describe scattering processes in which a new
electron-hole pair is created without changing the states of the
originally existing particles. Such processes are obviously strongly
off-resonant and will be disregarded.

For the two-particle terms (the two-particle part of the
electron-electron interaction) the Hamiltonian can be written as 
\begin{eqnarray}
 H^{(2)}&=&\frac{1}{2}\sum_{\gamma_1\gamma_2\gamma_3\gamma_4} V_{\gamma_1 \gamma_2 \gamma_3 \gamma_4}^{(2)}a_{\gamma_1}^{\dagger}a_{\gamma_2}^{\dagger}a_{\gamma_3}a_{\gamma_4},
\end{eqnarray}
where 
\begin{eqnarray}\label{matrix_elements1}
V_{\gamma_1\gamma_2\gamma_3\gamma_4}&=&\sum_{s,s'}\int d^3r\int d^3r'\Psi_{\gamma_1}^{\star}(\bm{r},s)\Psi_{\gamma_2}^{\star}(\bm{r}',s')\nonumber\\
& &\times U^{(2)}(\bm{r},\bm{r}')\Psi_{\gamma_3}(\bm{r}',s')\Psi_{\gamma_4}(\bm{r},s),
\end{eqnarray}
and $U^{(2)}(\bm{r},\bm{r}')$ represents all the two-particle terms of the Coulomb interaction.
Hence, the matrix elements are 
\begin{eqnarray}
 & &\langle\!\alpha_1\alpha_2\beta_1\beta_2|H^{(2)}|\alpha\beta\rangle\!=\frac{1}{2}\sum_{\gamma_1\gamma_2\gamma_3\gamma_4}V_{\gamma_1\gamma_2\gamma_3\gamma_4}^{(2)}
 \nonumber\\
 & &\times \langle\!GS|a_{\beta_2}^{\dagger}a_{\beta_1}^{\dagger}a_{\alpha_2}a_{\alpha_1}
 a_{\gamma_1}^{\dagger}a_{\gamma_2}^{\dagger}a_{\gamma_3}a_{\gamma_4}a_{\alpha}^{\dagger}a_{\beta}|GS\rangle\!.
\end{eqnarray}

There are four assignments of the indices $\gamma_1\gamma_2\gamma_3\gamma_4$ to 
the valence ($v$) and conduction ($c$) bands that lead to non-zero
matrix elements: (A)~vcvv, (B)~cvvv, (C)~ccvc, and (D)~cccv.

For these assignments the matrix elements are 
\begin{widetext}
\begin{eqnarray}
 \lefteqn{\langle\!\alpha_1\alpha_2\beta_1\beta_2|H^{(2)}|\alpha\beta\rangle_A=\langle\!\alpha_1\alpha_2\beta_1\beta_2|H^{(2)}|\alpha\beta\rangle_B}\nonumber\\
 &=&\frac{1}{2}\sum_{\beta'}\left[
 -V_{\beta'\alpha_1\beta_1\beta'}\delta_{\beta\beta_2}\delta_{\alpha\alpha_2}+
  V_{\beta'\alpha_2\beta_1\beta'}\delta_{\beta\beta_2}\delta_{\alpha\alpha_1}+
  V_{\beta'\alpha_1\beta_2\beta'}\delta_{\alpha\alpha_2}\delta_{\beta\beta_1}-
  V_{\beta'\alpha_2\beta_2\beta'}\delta_{\alpha_1\alpha}\delta_{\beta\beta_1}
  \right]\nonumber\\
 & &+\frac{1}{2}\sum_{\beta'}\left[
 V_{\beta'\alpha_1\beta'\beta_1}\delta_{\beta\beta_2}\delta_{\alpha\alpha_2}-
 V_{\beta'\alpha_1\beta'\beta_2}\delta_{\beta\beta_1}\delta_{\alpha\alpha_2}-
 V_{\beta'\alpha_2\beta'\beta_1}\delta_{\beta\beta_2}\delta_{\alpha\alpha_1}+
 V_{\beta’\alpha_2\beta’\beta_2}\delta_{\beta\beta_1}\delta_{\alpha\alpha_1} \right]\nonumber\\
 & &+\frac{1}{2}\left[
 V_{\beta\alpha_1\beta_1\beta_2}\delta_{\alpha\alpha_2}-
 V_{\beta\alpha_2\beta_1\beta_2}\delta_{\alpha\alpha_1}-
 V_{\beta\alpha_1\beta_2\beta_1}\delta_{\alpha\alpha_2}+
 V_{\beta\alpha_2\beta_2\beta_1}\delta_{\alpha\alpha_1} \right],\label{AB}
\end{eqnarray}
and
\begin{eqnarray}
 \lefteqn{\langle\!\alpha_1\alpha_2\beta_1\beta_2|H^{(2)}|\alpha\beta\rangle_C=\langle\!\alpha_1\alpha_2\beta_1\beta_2|H^{(2)}|\alpha\beta\rangle_D}\nonumber\\
 &=&\frac{1}{2}\left[
-V_{\alpha_1\alpha_2\beta_1\alpha}\delta_{\beta\beta_2}+
 V_{\alpha_2\alpha_1\beta_1\alpha}\delta_{\beta\beta_2}+
 V_{\alpha_1\alpha_2\beta_2\alpha}\delta_{\beta\beta_1}-
 V_{\alpha_2\alpha_1\beta_2\alpha}\delta_{\beta\beta_1} \right],\label{CD}
\end{eqnarray}
\end{widetext}
where we used the symmetry $V_{\gamma_1\gamma_2\gamma_3\gamma_4}=V_{\gamma_2\gamma_1\gamma_4\gamma_3}$.

The first two lines in the contributions $A$ and $B$ contain the direct and
exchange interactions with all the other electrons in the NC. 
For instance, the first term in Eq.~\eqref{AB} describes a direct 
Coulomb process in which an electron in the valence band state $\beta_1$ 
scatters off all the electrons in the valence band and makes a 
transition to a conduction band state $\alpha_1$. The following 
three terms account for the same process but with the initial 
state $\beta_2$ and final state $\alpha_2$. These direct terms 
cancel the electron-ion interactions in the leading order (on the mesoscopic scale).
Similarly, the four terms in the second line of Eq.~\eqref{AB} describe 
exchange scattering of valence band electrons off all the other electrons 
with a transition to the conduction band.
These exchange terms are not so straightforward to treat as the direct ones. However, all 
the terms containing two Kronecker deltas 
like $\delta_{\alpha\alpha_i}\delta_{\beta\beta_j}$, 
couple the two-particles state $|\alpha\beta\rangle$ 
to a four-particle state with two particles 
(electron and hole) in the same state; like 
$|\alpha\alpha_2,\beta\beta_2\rangle$. 
These two states differ considerably by energy 
(two particles do not change their states but a 
new e-h pair is created), hence these terms describe strongly
off-resonant couplings and can be neglected.
The last lines in the contributions $A$ and $B$ describe scattering processes in which 
one electron is a spectator, while a hole 
changes its state and induces generation of the second e-h pair.
Since the hole energies are typically smaller, these processes are of relatively little importance. 
Therefore, for the further calculations, we are left with the (identical)
terms $C$ and $D$ that describe scattering processes in which an electron
makes and intraband transition and transfer its energy to an inter-band
excitation that produces another e-h pair.

\subsection{Matrix elements}
\label{sec:Matrix elements}
In this section we calculate the matrix elements for the explicit model 
of wave functions defined in Sec.~\ref{sec:Single particle states}. This is done by expressing the 
result in terms of standard (envelope-function) Coulomb integrals and the 
single-band interband matrix elements found previously\cite{kowalski09}.

The matrix elements are written as in Eq.~\eqref{matrix_elements1}


Upon substitution of Eq.~\eqref{wavefunction} in Eq.~\eqref{matrix_elements1} one has
\begin{widetext}
\begin{eqnarray*}
V_{\gamma_1\gamma_2\gamma_3\gamma_4}&=&
\sum_{\lambda_1\lambda_2\lambda_3\lambda_4}\sum_{\nu_1\nu_2\nu_3\nu_4}
c_{\nu_1}^{{\star}(\gamma_1\lambda_1)}c_{\nu_2}^{{\star}(\gamma_2\lambda_2)}
c_{\nu_3}^{(\gamma_3\lambda_3)}c_{\nu_4}^{(\gamma_4\lambda_4)}
\sum_{\bm{R}\bm{R}'}\sum_{ss'}\int d^3\zeta \int d^3\zeta' \\
& & \times\psi_{\nu_1}^{\star}(\bm{R})u_{\lambda_1}^{\star}(\bm{\zeta} , s)\psi_{\nu_2}^{\star}(\bm{R}')u_{\lambda_2}^{\star}(\bm{\zeta}' , s') U^{(2)}(\bm{R}+\bm{\zeta},\bm{R}'+\bm{\zeta}')\psi_{\nu_3}(\bm{R}')u_{\lambda_3}(\bm{\zeta}' , s')\psi_{\nu_4}(\bm{R})u_{\lambda_4}(\bm{\zeta} , s)
\end{eqnarray*}
where we have followed the standard procedure of replacing the spatial integrals by summation over unit cells ($\bm{R}$) and integration over a
single unit cell ($\zeta$).
In view of orthogonality of Bloch functions,
\begin{equation}\label{orthogonality}
 \sum_{s}\int d^3\zeta u_{\lambda}^{\star}(\bm{\zeta} ,
 s)u_{\lambda'}(\bm{\zeta} , s)
=\delta_{\lambda \lambda'}v,\nonumber
\end{equation}
where, $v$ is the volume of the unit cell, 
two essentially different cases appear depending on the bands involved. 
If $\lambda_1=\lambda_4$ and $\lambda_2=\lambda_3$ then, in the leading order, 
one can set $U^{(2)}(\bm{R}+\bm{\zeta},\bm{R}'+\bm{\zeta}')=U^{(2)}(\bm{R},\bm{R}')$.
The corresponding contribution to the matrix element is
\begin{eqnarray*}
V_{\gamma_1\gamma_2\gamma_3\gamma_4}^{(0)}&=&
\sum_{\lambda_1\lambda_2}\sum_{\nu_1\nu_2\nu_3\nu_4}c_{\nu_1}^{{\star}(\gamma_1\lambda_1)}
c_{\nu_2}^{{\star}(\gamma_2\lambda_2)}c_{\nu_3}^{(\gamma_3\lambda_2)}c_{\nu_4}^{(\gamma_4\lambda_1)} 
\sum_{\bm{R}\bm{R}'}\psi_{\nu_1}^{\star}(\bm{R})\psi_{\nu_2}^{\star}(\bm{R}')
U^{(2)}(\bm{R},\bm{R}')\psi_{\nu_3}(\bm{R}')\psi_{\nu_4}(\bm{R}) \\
& &\times \sum_{s}\int d^3\zeta u_{\lambda_1}^{\star}(\bm{\zeta} , s)u_{\lambda_1}(\bm{\zeta} , s)\sum_{s'}
    \int d^3\zeta'u_{\lambda_2}^{\star}(\bm{\zeta}' , s')u_{\lambda_2}(\bm{\zeta}' , s').
\end{eqnarray*}

Using Eq.~(\ref{orthogonality}), and returning to integration according to $v\sum_{\bm{R}}\to \int d^3R$, 
one finds
\begin{equation*}
V_{\gamma_1\gamma_2\gamma_3\gamma_4}^{(0)}=\sum_{\lambda_1\lambda_2}\sum_{\nu_1\nu_2\nu_3\nu_4}
c_{\nu_1}^{{\star}(\gamma_1\lambda_1)}c_{\nu_2}^{{\star}
(\gamma_2\lambda_2)}c_{\nu_3}^{(\gamma_3\lambda_2)}c_{\nu_4}^{(\gamma_4\lambda_1)}
h_{\nu_1\nu_2\nu_3\nu_4}^{(0)}
\end{equation*}
where
\begin{equation}\label{h(0)}
h_{\nu_1\nu_2\nu_3\nu_4}^{(0)}=\int d^3R\int d^3R'\psi_{\nu_1}^{\star}(\bm{R}) \psi_{\nu_2}^{\star}(\bm{R}')U^{(2)}(\bm{R},\bm{R}')\psi_{\nu_3}(\bm{R}')\psi_{\nu_4}(\bm{R}).
\end{equation}

If $\lambda_1=\lambda_4$ but $\lambda_2\ne\lambda_3$ then the previously calculated contribution vanishes due to orthogonality of Bloch functions.
In this case, we expand 
\begin{eqnarray}
 U^{(2)}(\bm{R}+\bm{\zeta},\bm{R}'+\bm{\zeta}')\approx U^{(2)}(\bm{R},\bm{R}')+\bigtriangledown_{\bm{R}'}U^{(2)}(\bm{R},\bm{R}')\cdot\bm{\zeta}'\nonumber.
\end{eqnarray}
The corresponding contribution to the matrix element is then
\begin{eqnarray*}
V_{\gamma_1\gamma_2\gamma_3\gamma_4}^{(1a)}&=&
\sum_{\lambda_1\lambda_2\lambda_3}\sum_{\nu_1\nu_2\nu_3\nu_4}c_{\nu_1}^{{\star}(\gamma_1\lambda_1)}c_{\nu_2}^{{\star}(\gamma_2\lambda_2)}c_{\nu_3}^{(\gamma_3\lambda_3)}c_{\nu_4}^{(\gamma_4\lambda_1)} 
\sum_{\bm{R}\bm{R}'}\psi_{\nu_1}^{\star}(\bm{R})\psi_{\nu_2}^{\star}(\bm{R}')\bigtriangledown_{\bm{R}'}U^{(2)}(\bm{R},\bm{R}')\psi_{\nu_3}(\bm{R}')\psi_{\nu_4}(\bm{R})\nonumber\\
& & \cdot \sum_{s}\int d^3\zeta u_{\lambda_1}^{\star}(\bm{\zeta} , s)u_{\lambda_1}(\bm{\zeta} , s)\sum_{s'}
    \int d^3\zeta'u_{\lambda_2}^{\star}(\bm{\zeta}' , s')\bm{\zeta}'u_{\lambda_3}(\bm{\zeta}' , s')\nonumber\\
&=&\sum_{\lambda_1\lambda_2\lambda_3}\sum_{\nu_1\nu_2\nu_3\nu_4}c_{\nu_1}^{{\star}(\gamma_1\lambda_1)}c_{\nu_2}^{{\star}(\gamma_2\lambda_2)}c_{\nu_3}^{(\gamma_3\lambda_3)}c_{\nu_4}^{(\gamma_4\lambda_1)}
h_{\nu_1\nu_2\nu_3\nu_4}^{(\lambda_1\lambda_2\lambda_3\lambda_1)},
\end{eqnarray*}
\end{widetext}
where, replacing summation over unit cells by integration as previously, 
\begin{eqnarray}
& &h_{\nu_1\nu_2\nu_3\nu_4}^{(\lambda_1\lambda_2\lambda_3\lambda_1)}=\int d^3R\int d^3R'\\
& &\times\psi_{\nu_1}^{\star}(\bm{R}) \psi_{\nu_2}^{\star}(\bm{R}')\bm{r}_{\lambda_2\lambda_3}\cdot\bigtriangledown_{\bm{R}'}U^{(2)}(\bm{R},\bm{R}')\psi_{\nu_3}(\bm{R}')\psi_{\nu_4}(\bm{R})\nonumber
\end{eqnarray}
and 
\begin{eqnarray}\label{rlambda1lambda2}
\bm{r}_{\lambda_2\lambda_3}=\frac{1}{V}\sum_{s}\int d^3\zeta' u_{\lambda_2}^{\star}(\bm{\zeta}' , s')\bm{\zeta}'u_{\lambda_3}(\bm{\zeta}' , s').\nonumber
\end{eqnarray}
In a similar way, if $\lambda_2=\lambda_3$ and $\lambda_1\ne\lambda_4$ then
\begin{eqnarray*}
\lefteqn{V_{\gamma_1\gamma_2\gamma_3\gamma_4}^{(1b)}=}\\
& &\sum_{\lambda_1\lambda_2\lambda_4}\sum_{\nu_1\nu_2\nu_3\nu_4}c_{\nu_1}^{{\star}(\gamma_1\lambda_1)}c_{\nu_2}^{{\star}(\gamma_2\lambda_2)}c_{\nu_3}^{(\gamma_3\lambda_2)}c_{\nu_4}^{(\gamma_4\lambda_4)}
h_{\nu_1\nu_2\nu_3\nu_4}^{(\lambda_1\lambda_2\lambda_2\lambda_4)},
\end{eqnarray*}
where
\begin{eqnarray}
h_{\nu_1\nu_2\nu_3\nu_4}^{(\lambda_1\lambda_2\lambda_2\lambda_4)}
=h_{\nu_2\nu_1\nu_4\nu_3}^{(\lambda_2\lambda_1\lambda_1\lambda_3)}.
\end{eqnarray}
The terms with $\lambda_1 \neq \lambda_4$ and $\lambda_2 \neq \lambda_3$ 
contribute only in the second order in the expansion of the 
Coulomb potential, hence are formally on the order of $(a/R)^2$ 
and will not be considered here. Thus, finally, one finds
\begin{eqnarray*}
V_{\gamma_1\gamma_2\gamma_3\gamma_4}&=&
\sum_{\lambda_1\lambda_2\lambda_3\lambda_4}\sum_{\nu_1\nu_2\nu_3\nu_4}
c_{\nu_1}^{{\star}(\gamma_1\lambda_1)}c_{\nu_2}^{{\star}(\gamma_2\lambda_2)}
c_{\nu_3}^{(\gamma_3\lambda_3)}c_{\nu_4}^{(\gamma_4\lambda_4)}
\\
& &\times 
\left\{
\begin{array}{ll}
h_{\nu_1\nu_2\nu_3\nu_4}^{(0)} & \text{if } \lambda_1=\lambda_4 \ \text{and}\ \lambda_2=\lambda_3,\\
h_{\nu_1\nu_2\nu_3\nu_4}^{(\lambda_1\lambda_2\lambda_3\lambda_1)} & 
\text{if } \lambda_1=\lambda_4 \ \text{and}\ \lambda_2\ne\lambda_3,\\
h_{\nu_2\nu_1\nu_4\nu_3}^{(\lambda_2\lambda_1\lambda_4\lambda_2)} & 
\text{if } \lambda_1\ne\lambda_4 \ \text{and}\ \lambda_2=\lambda_3,\\
0\ \ \ \ \ \ \ \ \ \ \  & 
\text{if } \lambda_1\ne\lambda_4 \ \text{and}\ \lambda_2\ne\lambda_3,\\
\end{array} \right.
\end{eqnarray*}
which includes terms up to the first order in the expansion of the Coulomb potential.
In this way, we have reduced the calculation of inter-band Coulomb matrix elements 
between X and BX states with 8-band wave functions to single-subband 
terms given by Eq.~(\ref{h(0)}) and Eq.~(\ref{rlambda1lambda2}). 
The former only involve the envelope functions that describe the carrier states 
on the mesoscopic level and can be calculated in a standard way. The latter depend 
on the microscopic, atomic-scale structure via the interband matrix element of the 
position vector $\bm{r}_{\lambda \lambda '}$, which is proportional to the inter-band 
dipole moment (involved in the optical selection rules). These microscopic terms for 
the basis states used here have been calculated in Ref.~[23].
Note that in the single-band approximation, when the states above and below the gap are assumed to 
be composed purely of the bulk conduction and valence band states, respectively, the 
mesoscopic term vanishes due to Bloch function orthogonality. In general,
due to band mixing in a strongly confining nanostructure, both the mesoscopic 
and microscopic contributions can be non-zero.

\section{Results}
\label{sec:results}
In this section, we present results of calculations performed within the eight-band 
model presented above. We focus on the comparison between the typical
magnitudes of the microscopic and mesoscopic contributions of the
inter-band Coulomb couplings and on the general statistical distribution of 
the coupling strengths between optically active (bright) X states and BX states 
vs. the energy difference between the two coupled states. 
In view of the enormous number of exciton and, in particular, biexciton states, 
the statistics of interband Coulomb couplings to be presented 
here are obtained by randomly selecting states from a broad energy 
range (using a uniform distribution over the set of quantum numbers).

\begin{figure}[tb]
\includegraphics[width=85mm]{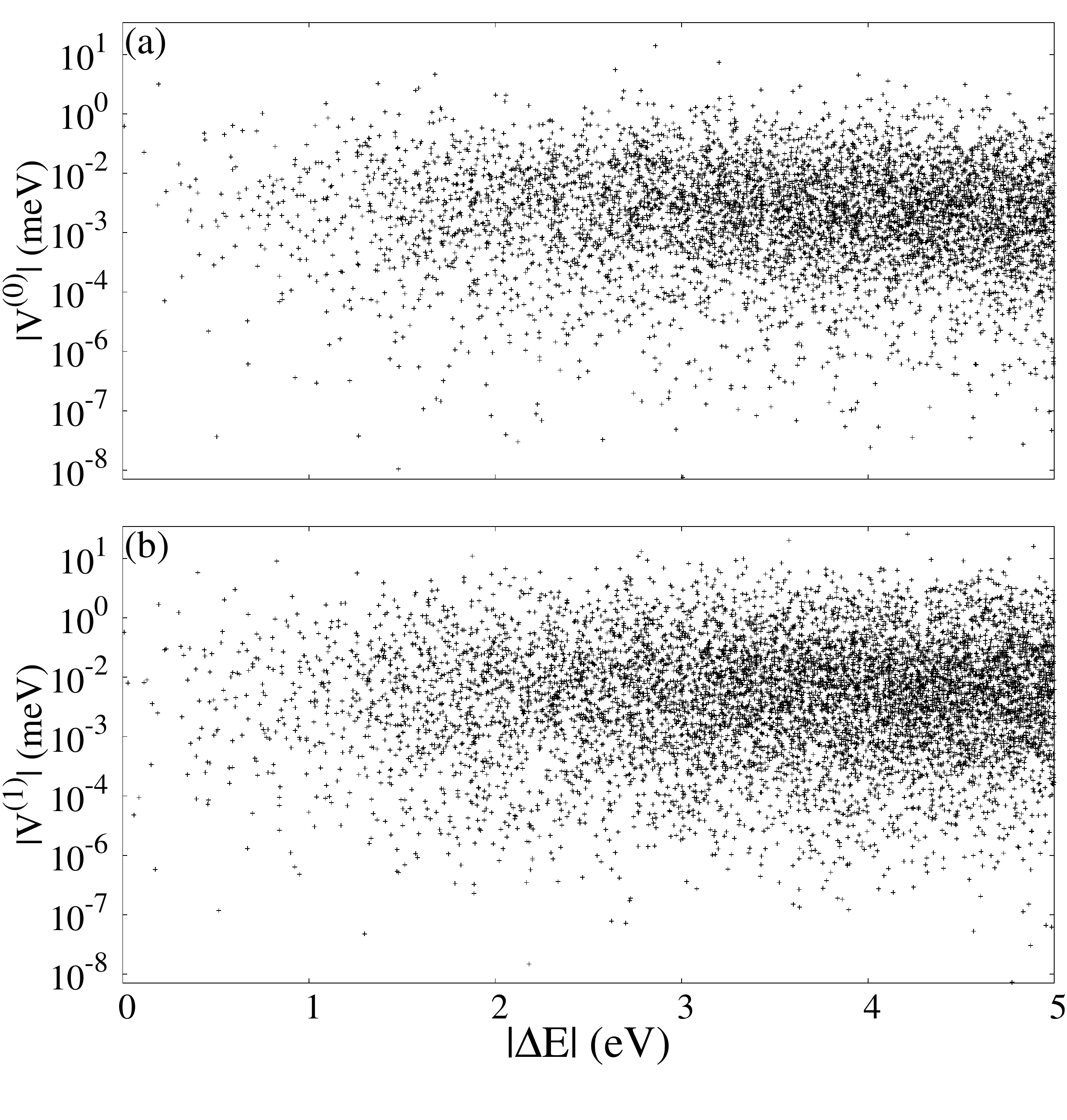}
\caption{The absolute value of the mesoscopic (a) and microscopic (b) contributions to the Coulomb 
coupling matrix elements between X and BX states vs. the 
energy distance between these states for a sample of $11\,000$ coupled
X-BX pairs.} 
\label{couplingDeltaEtot}
\end{figure}

In order to characterize the typical X-BX Coulomb coupling strengths and the 
distribution of the relative energies of coupled X-BX pairs, in 
Fig.~\ref{couplingDeltaEtot} we present the magnitudes of these couplings vs. the energy 
distance between the coupled states. The presented results are based on
about $50\,000$ randomly selected combinations of X and BX states with
the energy less  
than $5$ eV out of which $11\,000$ show non-zero coupling, 
which is still only a tiny fraction of the total number of possible X-BX
combinations. Each point corresponds to a single BX state  
coupled to an X state and its position shows the magnitude of the
Coulomb matrix element between these two states and the absolute value of the energy difference between 
these states.
The analysis is performed separately for the mesoscopic and microscopic contributions to the 
Coulomb matrix elements, $V^{(0)}$ and $V^{(1)}$, shown in 
Fig.~\ref{couplingDeltaEtot}(a) and Fig.~\ref{couplingDeltaEtot}(b), respectively. 
As can be seen in Fig.~\ref{couplingDeltaEtot}, the overall number of coupled pairs 
grows with increasing energy difference. This is due to rapid 
increase of the density of states of both X and BX states at higher energies.
Typical orders of magnitude for the X-BX coupling are up to several meV. Although we 
have found a small number of stronger couplings, about 100~meV, they only appear between energetically 
very distant states. Apart from this upper bound on the magnitudes 
of the matrix elements, Fig.~\ref{couplingDeltaEtot} indicates that matrix elements 
with values below $10^{-6}$~meV are unusual, which agrees with the earlier 
atomistic results for another material system\cite{korkusinski11}.

\begin{figure}[tb]
\includegraphics[width=85mm]{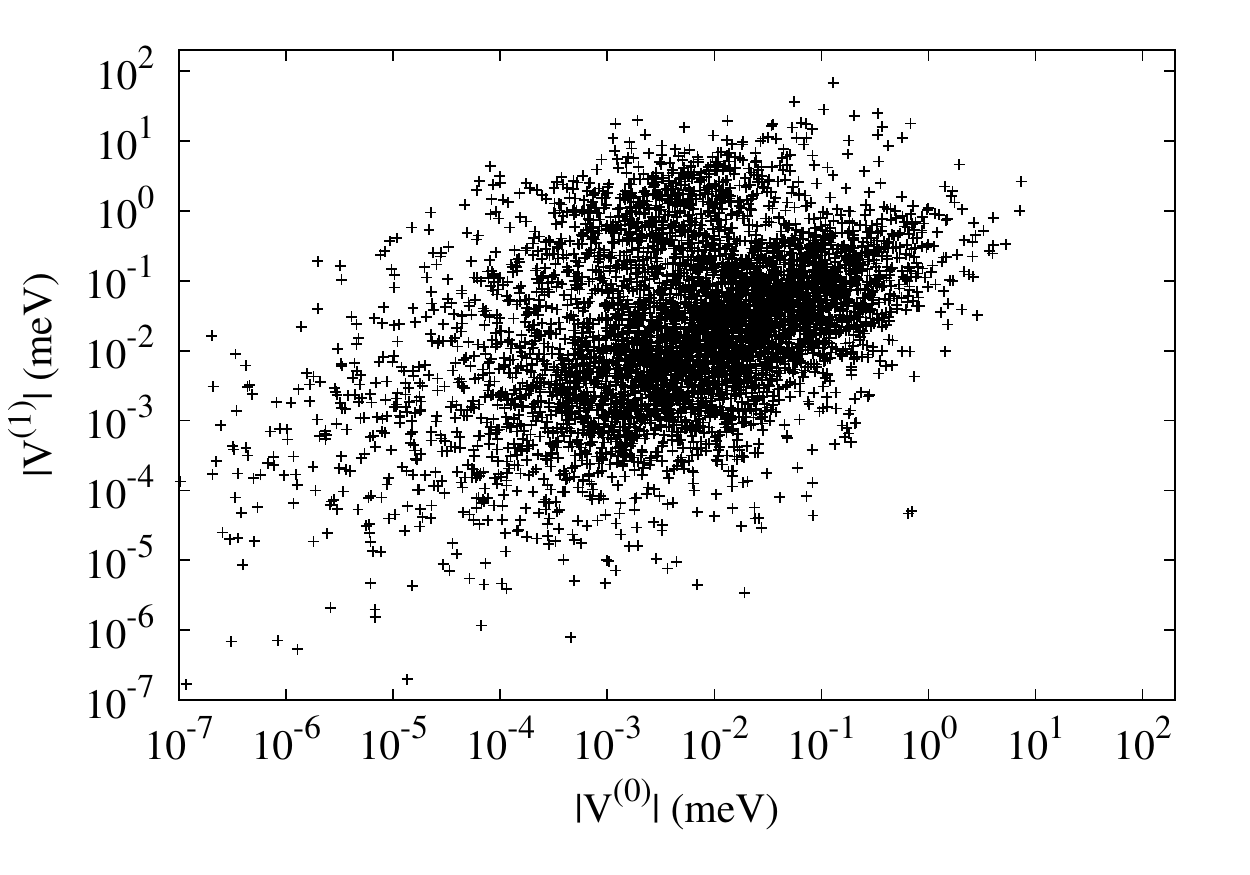}
\caption{Comparison of the magnitudes of micro- and mesoscopic
  contributions to the interband Coulomb coupling. Each point
  corresponds to one coupled $X-BX$ pair from a sample of $11\,000$
  and its position represents the magnitudes of the two contributions
  to the coupling.}  
\label{v0v1tot}
\end{figure}

From Fig.~\ref{couplingDeltaEtot} it is clear that the microscopic and mesoscopic contributions 
to the Coulomb coupling tend to be roughly of the same order of magnitude. 
This is confirmed in Fig.~\ref{v0v1tot}, where the magnitudes of the contributions to the  
Coulomb coupling for the same number of the combinations 
as in Fig.~\ref{couplingDeltaEtot} are compared. 
In vast majority of coupled X-BX configurations both contributions are non-zero, 
which results from identical selection rules for these two couplings 
(the fractions of cases with only $V^{(0)}$ or only $V^{(1)}$ non-zero are 
about 0.2\% and 4\%, respectively). As we can see, although the ratio of the two contributions 
in individual cases can vary over 10 orders of magnitude (roughly from $10^{-5}$ to $10^{5}$), 
in most cases they are almost of the same order of magnitude.

\begin{figure}[tb]
\includegraphics[width=85mm]{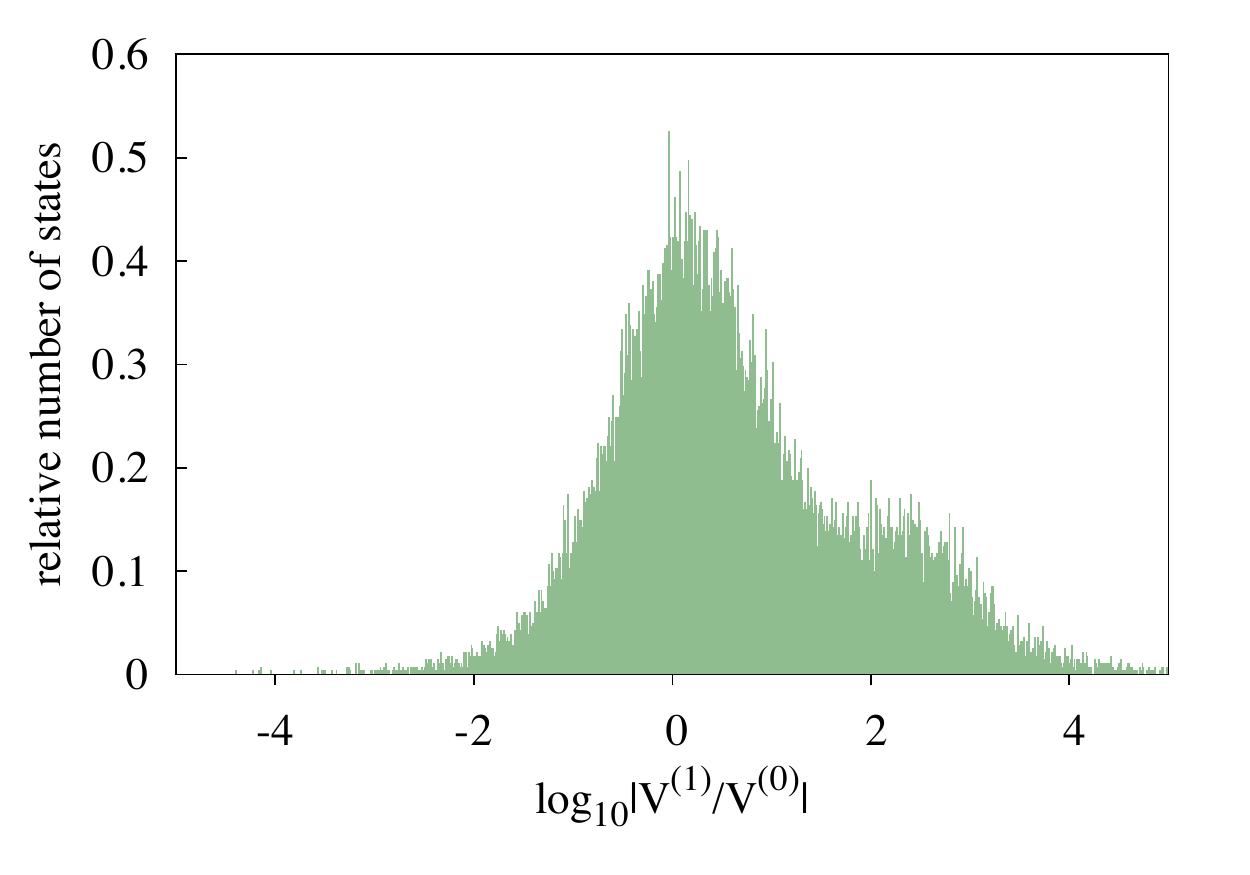}
\caption{Normalized distribution of the relative 
magnitudes of the micro- and mesoscopic couplings based on a sample of
about $40\,000$ coupled X-BX pairs.} 
\label{hist-article}
\end{figure}

The same property can be seen when one looks at the histogram showing 
the number of state combinations as a function of
$\log|V^{(1)}/V^{(0)}|$ (Fig.~\ref{hist-article}). Here we used a
larger sample of $200\,000$ X-BX pairs out of which over $40\,000$
were coupled.
The bimodal form of the distribution reflects the 
two groups of points visible in Fig.~\ref{hist-article}, corresponding to the cases where 
the two contributions are of similar magnitude and those where the microscopic 
contribution dominates by about two orders of magnitude. 
The origin of this special distribution remains unclear.

\begin{figure}[tb]
\includegraphics[width=85mm]{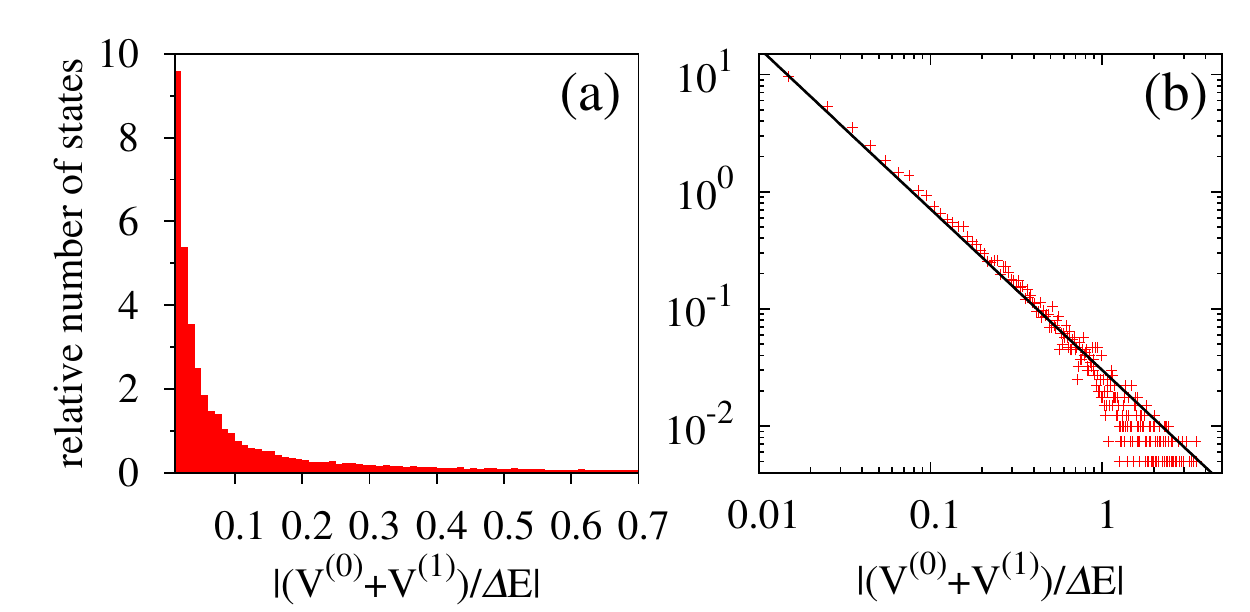}
\caption{Histogram of the values of the ratio
  $q=|(V^{(0)}+V^{(1)})/\Delta E|$ for a sample of
about $40\,000$ coupled X-BX pairs in linear (a) and logarithmic (b)
scale. The solid line in (b) shows a $q^{-2}$ dependence.} 
\label{v1+v0}
\end{figure}
As discussed in Ref.~[23],
 the distribution of the ratio of the coupling magnitude 
$V=V^{(0)}+V^{(1)}$ to the energy separation between the coupled states $\Delta E$ 
is of major importance for the convergence of numerical procedures: Since q=$|V/\Delta E|$ 
determines (via perturbation theory) the admixture of a biexciton state to 
the optically active single-exciton state, the distribution of this quantity 
must be integrable at $q\to 0$ in order for the computations to be convergent with respect 
to the width of the energy window in which the states have been found 
(which is always limited by the available computational resources). 
Statistics based on a simplistic model of carrier states yielded a $q^{-2}$ 
dependence as $q\to 0$, which provides a bound for the contribution of 
remote states (in view of the bounded values of the coupling magnitudes) 
and thus assures convergence\cite{kowalski09}. 
The results presented in Fig.~\ref{v1+v0} (based on the sample as in
Fig.~\ref{hist-article}) confirm that the same $q^{-2}$ 
form of the distribution is found in the present, more realistic
model, as shown by the solid line in Fig.~\ref{v1+v0}(b).

\section{Conclusions}
\label{sec:concl}
We have presented a method for calculating Coulomb matrix elements 
between exciton and biexciton states in a spherical semiconductor 
nanoparticle within the multi-band envelope function formalism based 
on the carrier states resulting from diagonalization of the 8-band 
\mbox{$\bm{k}\cdot\bm{p}$} Hamiltonian. We have shown that this coupling 
includes two contributions of different form: The mesoscopic one relies 
on the usual inter-band Coulomb matrix elements that contribute to inter-band 
couplings via band mixing. The microscopic contribution involves Coulomb 
matrix elements at the level of Bloch functions and does not vanish 
even if band mixing is neglected.

The relatively low computational cost of the \mbox{$\bm{k}\cdot\bm{p}$} 
method allowed us to build statistics of the coupling values over $\sim 10^5$ 
X-BX pairs in a broad energy window relevant, e.g., to photoelectric cell
operation. We have shown that the relative magnitude of these two contributions 
over a large statistical sample of X-BX pairs has a bimodal distribution with 
either both contributions equal or the microscopic one dominating roughly by 
two orders of magnitude.

We have also shown within our multi-band model that the ratio of 
the coupling magnitude to the energy separation between the coupled 
states follows a power-law distribution the exponent of which guarantees 
convergence of numerical calculations with respect to the width of the 
energy window to which such a computation must always be limited.

\begin{center}
\normalsize \textbf{Acknowledgments}
\end{center}
This work was supported by the TEAM programme of the Foundation 
for Polish Science co-financed from the European Regional Development Fund.


\begin{thebibliography}{10}

\bibitem{talapin10}
D.~V. Talapin, J.-S. Lee, M.~V. Kovalenko, and E.~V. Shevchenko, Chem. Rev.
  {\bf 110},  389  (2010).

\bibitem{teperik12}
T.~V. Teperik and A. Degiron, Phys. Rev. Lett. {\bf 108},  147401  (2012).

\bibitem{LinPe}
L. Peng, L. Hu, and X. Fang, Advanced Materials {\bf 25},  5321  (2013).

\bibitem{binks11}
D.~J. Binks, Phys. Chem. Chem. Phys. {\bf 13},  12693  (2011).

\bibitem{deuk11}
K. Hyeon-Deuk and O.~V. Prezhdo, Nano Lett. {\bf 11},  1845  (2011).

\bibitem{deuk12}
K. Hyeon-Deuk and O.~V. Prezhdo, ACS Nano {\bf 6},  1239  (2012).

\bibitem{franceschetti06}
A. Franceschetti, J.~M. An, and A. Zunger, Nano Lett. {\bf 6},  2191  (2006).

\bibitem{rabani08}
E. Rabani and R. Baer, Nano Lett. {\bf 8},  4488  (2008).

\bibitem{califano09}
M. Califano, ACS Nano {\bf 3},  2706  (2009).

\bibitem{baer12}
R. Baer and E. Rabani, Nano Lett. {\bf 12},  2123  (2012).

\bibitem{allan06}
G. Allan and C. Delerue, Phys. Rev. B {\bf 73},  205423  (2006).

\bibitem{delerue10}
C. Delerue, G. Allan, J.~J.~H. Pijpers, and M. Bonn, Phys. Rev. B {\bf 81},
  125306  (2010).

\bibitem{shabaev06}
A. Shabaev, A.~L. Efros, and A.~J. Nozik, Nano Lett. {\bf 6},  2856  (2006).

\bibitem{witzel10}
W.~M. Witzel, A. Shabaev, C.~S. Hellberg, V.~L. Jacobs, and A.~L. Efros, Phys.
  Rev. Lett. {\bf 105},  137401  (2010).

\bibitem{Silvestri}
L. Silvestri and V.~M. Agranovich, Phys. Rev. B {\bf 81},  205302  (2010).

\bibitem{korkusinski10}
M. Korkusinski, O. Voznyy, and P. Hawrylak, Phys. Rev. B {\bf 82},  245304
  (2010).

\bibitem{sambur10}
J.~B. Sambur, T. Novet, and B.~A. Parkinson, Science {\bf 330},  63  (2010).

\bibitem{semonin11}
O.~E. Semonin, J.~M. Luther, S. Choi, H.-Y. Chen, J. Gao, A.~J. Nozik, and
  M.~C. Beard, Science {\bf 334},  1530  (2011).

\bibitem{schulze11}
F. Schulze, M. Schoth, U. Woggon, A. Knorr, and C. Weber, Phys. Rev. B {\bf
  84},  125318  (2011).

\bibitem{maryam}
M. Azizi and P. Machnikowski, Phys. Rev. B {\bf 88},  115303  (2013).

\bibitem{schaller05}
R.~D. Schaller, V.~M. Agranovich, and V.~I. Klimov, Nat. Phys. {\bf 1},  189
  (2005).

\bibitem{rupasov07}
V.~I. Rupasov and V.~I. Klimov, Phys. Rev. B {\bf 76},  125321  (2007).

\bibitem{kowalski09}
P. Kowalski, {\L}. Marcinowski, and P. Machnikowski, Phys. Rev. B {\bf 87},
  075309  (2013).

\bibitem{korkusinski11}
M. Korkusinski, O. Voznyy, and P. Hawrylak, Phys. Rev. B {\bf 84},  155327
  (2011).

\bibitem{silvestri10}
L. Silvestri and V.~M. Agranovich, Phys. Rev. B {\bf 81},  205302  (2010).

\bibitem{piryatinski10}
A. Piryatinski and K.~A. Velizhanin, J. Chem. Phys. {\bf 133},  084508  (2010).

\bibitem{Velizhanin}
K.~A. Velizhanin and A. Piryatinski, Phys. Rev. Lett. {\bf 106},  207401
  (2011).

\bibitem{Kadantsev}
E. Kadantsev and P. Hawrylak, Phys. Rev. B {\bf 81},  045311  (2010).

\bibitem{efros98}
A.~L. Efros and M. Rosen, Phys. Rev. B {\bf 58},  7120  (1998).

\bibitem{brus84}
L.~E. Brus, J. Chem. Phys. {\bf 80},  4403  (1984).

\end{thebibliography}

\end{document}